\pdfoutput=1
\documentclass{article}

\usepackage[final]{neurips_2021}

\usepackage[utf8]{inputenc} 
\usepackage[T1]{fontenc}    
\usepackage[hidelinks]{hyperref} 
\usepackage{url}            
\usepackage{booktabs}       
\usepackage{amsfonts}       
\usepackage{nicefrac}       
\usepackage{microtype}      
\usepackage{xcolor}         
\usepackage[hang,flushmargin]{footmisc}
\usepackage{enumitem}
\usepackage{xspace}
\usepackage{arydshln}
\usepackage{multirow}
\usepackage{graphicx}

\usepackage{etoolbox}
\makeatletter
\patchcmd{\@verbatim}
  {\verbatim@font}
  {\verbatim@font\small}
  {}{}
\makeatother

\newcommand{\ignore}[1]{}
\newcommand{\mytt}[1]{\texttt{\small #1}}

\title{Building a Culture of Reproducibility\\ in Academic Research}

\author{Jimmy Lin \\[1ex]
David R. Cheriton School of Computer Science\\
University of Waterloo
}

\begin{document}

\maketitle

\begin{abstract}
Reproducibility is an ideal that no researcher would dispute ``in the abstract'', but when aspirations meet the cold hard reality of the academic grind, reproducibility often ``loses out''.
In this essay, I share some personal experiences grappling with how to operationalize reproducibility while balancing its demands against other priorities.
My research group has had some success building a ``culture of reproducibility'' over the past few years, which I attempt to distill into lessons learned and actionable advice, organized around answering three questions:\ why, what, and how.
I believe that reproducibility efforts should yield easy-to-use, well-packaged, and self-contained software artifacts that allow others to reproduce and generalize research findings.
At the core, my approach centers on self interest:\ I argue that the primary beneficiaries of reproducibility efforts are, in fact, those making the investments.
I believe that (unashamedly) appealing to self interest, augmented with expectations of reciprocity, increases the chances of success.
Building from repeatability, social processes and standardized tools comprise the two important additional ingredients that help achieve aspirational ideals.
The dogfood principle nicely ties these ideas together.
\end{abstract}

\section{Introduction}

I am passionate about making research reproducible by building and sharing, together with members of my research group and our collaborators around the world, software artifacts that others can use to recreate both our own work and the work of other researchers.
In my experience, reproducibility is an ideal that no researcher would dispute ``in the abstract'':\ like puppies, kittens, and rainbows, who could argue against it?
So why aren't we ``doing better'' when it comes to reproducibility?

To be fair, the state of affairs {\it has} much improved, compared to, say, a decade ago.
Today, it's fairly standard practice (and even expected?)\ in artificial intelligence and deep learning research that each paper is accompanied by a code repository that could (in theory) be used to reproduce the results.
Many researchers make model checkpoints publicly available (e.g., on the Huggingface model hub).
Yet, we are still falling short.
For example, \citet{Voorhees_etal_2016} examined 79 so-called ``Open Runs'' to TREC 2015---defined as a TREC submission self-identified as being backed by a software repository that can be used to recreate the exact run---and found that none of them were reproducible.
With improved tooling, one might expect the situation to be better today.
For example, computational notebooks are often touted as a solution to reproducibility because of their shareable and self-documenting nature.
However, in a large-scale study encompassing over 800k executions of publicly accessible valid Python notebooks on GitHub, \citet{Pimentel_etal_2019} found that only 24\% executed without errors and only 4\% produced the same results.

The truth is that while reproducibility is a noble goal, it is {\it aspirational}, not obligatory.
Unless, for example, publication venues enforce reproducibility (which would be very difficult to operationalize) competing priorities will take precedence.
Work on a new paper or create a reproduction package for an already published paper?
The choice is usually clear for most.
When aspirations meet the cold hard reality of the academic grind, it's almost inevitable that they lose.
Hence the (sad) state of reproducibility today.

The goal of this essay is to offer a possible path forward towards building a culture of reproducibility.
My approach can be summarized in these two high-level bits of advice:

\begin{itemize}[leftmargin=0.75cm]

\item Appeal to self interest instead of altruism.

\item Engineer social processes to promote virtuous cycles and build standardized tools to reduce technical barriers.

\end{itemize}

The first point tackles the incentive structure of ``why should I do this?''
Once that's been addressed, efforts should focus on promoting virtuous cycles and reducing technical barriers to make it easier to put reproducibility into practice.
In a way, both these points can be subsumed under the software development principle of ``eat your own dog food''.

Before proceeding any further, it is necessary to circumscribe the scope (and possible limitations) of the advice I'm offering.
At a high level, my research has been driven by the quest to develop techniques and build tools that connect users to relevant information.
Most of my focus has been on text, and so, cast into academic silos, my work lies at the intersection of natural language processing (NLP) and information retrieval (IR).
Nearly all the work from my research group can be characterized as applied and empirical in nature.
My latest interests lie in using pretrained transformer models to tackle text ranking and related challenges~\citep{Lin_etal_2021_ptr4tr}, particularly from the perspective of representation learning~\citep{Lin_arXiv2021_repir}.

Despite differences in research agendas, I {\it do} believe that much of my advice is applicable to empirical research across many sub-disciplines in computer science beyond NLP and IR, for example, computer vision and data mining.
Finally, it is worth noting that this essay represents an academic perspective:\ specifically, one of my most important roles is to mentor students.
Although much of what I write doesn't really apply to {\it employees} in a corporate context, some lessons can still be adapted.

This essay is organized around three main questions:

\begin{itemize}[leftmargin=0.75cm]

\item Why? (Section~\ref{section:why}) Why should I care and why should I do it?\footnote{The ``I'' in this case mostly refers to students. The real challenge is:\ How does an advisor convince students to actually do these things?}

\item What? (Section~\ref{section:what}) Okay, you've convinced me. But what does it actually mean to make research reproducible?

\item How? (Section~\ref{section:how}) Okay, you've shown me the end goal. But how do I get there?

\end{itemize}

Before concluding, Section~\ref{section:other} discusses a smörgåsbord of related issues.

One final preamble before getting underway.
To be precise, I use the term reproducibility in the sense articulated by the ACM in its Artifact Review and Badging Policy,\footnote{\url{https://www.acm.org/publications/policies/artifact-review-and-badging-current}} characterized as ``different team, same experimental setup''.
Specifically, ``this means that an independent group can obtain the same result using the author's own artifacts.''
In contrast, replicability can be characterized as ``different team, different experimental setup'', and operationally, ``this means that an independent group can obtain the same result using artifacts which they develop completely independently.''
Finally for completeness, repeatability can be characterized as ``same team, same experimental setup'', i.e., ``a researcher can reliably repeat her own computation.''
I use these terms in the way prescribed by this ACM policy.\footnote{A confusing detail:\ A previous version of the same ACM policy swapped the meaning of reproducibility and replicability.}

\section{Why?}
\label{section:why}

Broadly speaking, there are two main categories of arguments for why reproducibility is important and a worthwhile goal:\ the first is ``good science'' and the second is ``good citizenship''.
Both appeal to one's sense of altruism.

The first broad category of arguments, ``good science'', goes something like this:\ Science represents a systematic attempt to accumulate and organize knowledge about the world in the form of testable explanations and predictions (paraphrased from Wikipedia).
In the computational sciences, reproducibility and replicability are the mechanisms by which researchers can build on each other's results to accumulate knowledge.
Reproducible and replicable findings increase the veracity of the underlying scientific claims.
Conversely, the inability to reproduce or replicate a finding casts doubts over whether it can be reliably built on or extended.
If science is metaphorically the process of standing on the shoulder of giants in order to see further, reproducibility and replicability are the processes by which we test the stability of those shoulders we're attempting to climb on.
This is the general sentiment expressed in~\cite{Sonnenburg_etal_JMLR2012}, but see~\cite{Drummond_2009} for counterpoints.

The second broad category of arguments, ``good citizenship'', goes something like this:\ A large part of the funding for research is provided by various governments via tax dollars, and thus, it behooves researchers to share their results in the broadest way possible.
This usually entails making publications and associated research data publicly accessible---because, after all, it belongs to the people who ultimately supported the research (i.e., taxpayers).
In Canada, this is enshrined in policy:\ the tri-agencies, which are federal granting agencies that promote and support research, hold the official position ``that research data collected through the use of public funds should be responsibly and securely managed and be, where ethical, legal and commercial obligations allow, available for reuse by others.''\footnote{\url{https://www.science.gc.ca/eic/site/063.nsf/eng/h_97610.html}}
For research data management, the agencies support the guiding principles commonly known as FAIR, which stands for Findable, Accessible, Interoperable, and Reusable~\citep{WilkinsonMark_etal_2016}.
The U.S.\ National Science Foundation\footnote{\url{https://www.nsf.gov/bfa/dias/policy/dmp.jsp}} and the European Commission\footnote{\url{https://open-research-europe.ec.europa.eu/for-authors/data-guidelines}} hold similar positions.
While the original policies were formulated specifically with research data in mind, increasingly, software artifacts are given similar treatments~\citep{Lamprecht_etal_2020,BarkerMichelle_etal_2022}, and hence there is an increasing emphasis by funders (and by extension the researchers they support) on reproducibility within the broad umbrella of data management and stewardship.

While both categories of arguments are undeniably persuasive, the unfortunate downside is that they appeal primarily to the researcher's altruism.
Even in the relatively narrow cases where there are clear mandates and directives (e.g., from funding agencies), unless there is alignment with self interest, researchers will tend to do the minimum required.
When ``noble aspirations'' come into competition with the daily grind of the academic existence with its numerous other priorities, the (sad) truth is that reproducibility often takes a back seat.
When faced with the choice of working on a new paper or cleaning up a paper that's already been published for reproducibility purposes, what do you think a student would do?

To better help researchers prioritize competing demands, I offer another motive for reproducibility that instead appeals to self interest:

\begin{itemize}[leftmargin=0.75cm]

\item I say to a student:\ you're not doing reproducibility for others; you're doing it for yourself.

\item I say to myself:\ effort invested in reproducibility will help my research group iterate more rapidly and thus become more productive overall.

\end{itemize}

To be precise, according to the ACM Artifact Review and Badging Policy, ``self-reproducibility'' is more accurately called repeatability, so I will use this term in the discussions below.

In my nearly two decades as a faculty, I have had the following scenario happen countless times:\ A student is no longer able to recreate the experimental results obtained a short while ago.\footnote{Here, I'm not talking about ``noise'' that can be attributed to, for example, non-determinism in GPU execution. These are cases where something clearly ``worked'', but now doesn't.}
That is, the results are not repeatable.
Perhaps there was a bug in the original implementation?
A parameter setting that ``leaked'' the test data?
An ``oracle'' assumption that was later removed?
There are numerous reasons why this could be so.

If the results have not been made public (i.e., part of ongoing work that has not yet been published), then the student can (and should) ``start over''.
A common case is a rejected paper where reviewers suggested new experimental conditions (e.g., ablation study).
If the old experiments are not repeatable, then the implementation should be checked again carefully and {\it all} experiments should be run anew.
And if the previously observed gains have now disappeared, it just means that the innovation was never real to begin with.

However, more vexing is the situation where the results have already been reported in a published paper.
That is, a student cannot repeat an experiment that has already been ``enshrined'' in the literature.
What to do then?
A common scenario is that the student is trying to repeat the previously published experiment in order to perform follow-up work or to use those results as the basis of new research.
Another common scenario is when the student is integrating several threads of already published work into a thesis, and the additional experiments are critical to knitting the otherwise disparate threads together.
Whatever the case, proceeding without attempting to rectify the repeatability failure is scientifically dubious.
In empirical research, experimental results always need points of reference for meaningful comparison (e.g., baselines, ablations, etc.), and referencing results that cannot be recreated is just bad science.

At this point, I typically urge in the strongest possible way that the student first resolve the repeatability failure, which can, unfortunately, involve quite a bit of effort.
The original experiments may have been conducted months ago, and in the frantic dash to a paper deadline, the student may not have been meticulous taking notes in a lab notebook.\footnote{To students:\ You do have a lab notebook, right?}
So, the starting point of the repeatability effort may be a directory containing files with names like \mytt{config3-bugfix-trial2.yaml}, or worse yet, just a bunch of poorly named result files (created by complex command-line invocations that weren't properly recorded).

Thus, I explain to students the importance of repeatability today to prevent future frustration:\ the consequences can range from a missed opportunity for another paper to a graduation roadblock.
As the saying goes, ``future you will thank you!''
Note that, critically, the motivation for the student is self interest, {\it not} altruism.
Repeatability is a good first step towards reproducibility.
In fact, I would characterize the combination along these lines:
\begin{equation}
\textrm{repeatability} + \textrm{social processes} + \textrm{standardized tools} = \textrm{reproducibility}
\end{equation}
Given repeatability as a starting point, we can ``get to'' reproducibility by engineering social processes to promote virtuous cycles and building standardized tools to reduce technical barriers.
These I argue are the key ingredients to building a culture of reproducibility.
Section~\ref{section:how} explains these two elements in much more detail.

\section{What?}
\label{section:what}

Having covered the ``why'', I'll move on to cover the ``what''.
Specifically, I'll try to answer the question:\ What does ``good reproducibility'' look like?
That is, I'll start at the end by describing my ``aspirational ideal'' of what the end result of a reproducibility effort should be.

The broader context of this discussion is what information retrieval researchers call the ``core'' ranking problem (also called {\it ad hoc} retrieval).
I'll just wholesale lift the definition from~\cite{Lin_arXiv2021_repir}:\
Given an information need expressed as a query $q$, the task is to return a ranked list of $k$ documents\footnote{Consistent with parlance in information retrieval, I use ``document'' in a generic sense to refer to the unit of retrieved text, even though in truth it may be a passage, a web page, a PDF, or some arbitrary span of text.} $\{d_1, d_2 \ldots d_k\}$ from an arbitrarily large but finite collection of documents $\mathcal{D} = \{ d_i \}$ that maximizes a metric of interest, for example, nDCG, AP, etc.
These metrics vary, but they all aim to quantify the ``goodness'' of the results with respect to the information need.
The retrieval task is also called top-$k$ retrieval (or ranking), where $k$ is the length of the ranked list.

Generically, a retrieval model is a software artifact that addresses the core ranking problem, and one main focus of many researchers is to build better such models.
This is primarily an empirical endeavor, as the dominant way of demonstrating (i.e., in academic publications) that one model is better than another is based on measurements using test collections.

In the context of retrieval (or ranking) models, I believe that reproducibility efforts should yield easy-to-use, well-packaged, and self-contained software artifacts with clearly defined scope that allow the broadest possible audience to reproduce research findings by recreating experimental results.
Ideally, the artifact should support what I call ``two-click reproductions'', where a user can reproduce a reported result (for example, from a paper) with only two clicks:\ one click to copy a command-line invocation from a source (for example, a documentation page) and another click to paste the command into a shell.
How to ensure that the two-click reproductions ``work as advertised'' will be discussed in Section~\ref{section:how}.

I believe that my research group---with the help of external collaborators---has achieved this ``aspirational ideal'' in many parts of our two IR toolkits, Anserini~\citep{Yang_etal_SIGIR2017,Yang_etal_JDIQ2018} and Pyserini~\citep{Lin_etal_SIGIR2021_Pyserini}.
Anserini is built around the open-source Lucene search library, the most widely adopted solution for tackling search problems in deployed real-world applications (typically, via platforms such as Elasticsearch).
Our goal in building a {\it research} toolkit around Lucene is to facilitate a two-way exchange between academia and industry~\citep{Devins_etal_WSDM2022}.
Pyserini provides Python bindings to the capabilities offered in Anserini and integration with neural retrieval models built on industry-standard packages such as Huggingface Transformers, PyTorch, and Faiss.
Pyserini includes many dense and sparse retrieval models built on transformer-based encoders as well as traditional ``bag-of-words'' models such as BM25 and relevance feedback techniques.

To provide a specific example with Pyserini, experiments applying our uniCOIL model~\citep{Lin_Ma_arXiv2021} to the development queries of the MS MARCO passage ranking task~\citep{MS_MARCO_v3} can be accomplished with the following command:
\begin{quote}
\begin{verbatim}
python -m pyserini.search.lucene \
  --index msmarco-passage-unicoil-d2q \
  --topics msmarco-passage-dev-subset \
  --encoder castorini/unicoil-msmarco-passage \
  --output runs/run.msmarco-passage.unicoil.tsv \
  --output-format msmarco \
  --batch 36 --threads 12 \
  --hits 1000 \
  --impact
\end{verbatim}
\end{quote}
\noindent In terms of ease of use, a researcher who wishes to reproduce the results reported in our paper can do so with a single command, via two clicks:\ copy and paste (i.e., ``two-click reproduction'').
The above command calls the main driver program \mytt{pyserini.search.lucene} in a package that is published in the Python Package Index (PyPI),\footnote{\url{https://pypi.org/project/pyserini/}} thus making the software artifact well-packaged, since it can be installed with standard tools such as \mytt{pip}.

The two-click reproduction described above is self-contained because it has no other dependencies---many details are handled ``behind the scenes''.
For example:

\begin{itemize}[leftmargin=0.75cm]

\item The option \mytt{{-}{-}index} specifies an inverted index of a commonly used corpus in IR research (the MS MARCO passage corpus) that Pyserini already ``knows about'', along with dozens of other common corpora.
On first invocation, Pyserini downloads a copy of the index from a known location (servers at the University of Waterloo) and caches it locally.

\item The option \mytt{{-}{-}topics} specifies a standard set of queries that is already included in Pyserini, so the user doesn't need to visit a separate website to download them.

\item The option \mytt{{-}{-}encoder} refers to a transformer model for encoding the queries, which is hosted on the Huggingface model hub; Pyserini downloads and caches the model locally.

\end{itemize}

The execution of the above command yields a run file in the MS MARCO document format that can then be fed into the official MS MARCO scoring script to arrive at the official evaluation metric, reciprocal rank at cutoff 10.
This scoring script is also conveniently packaged in Pyserini:
\begin{quote}
\begin{verbatim}
python -m pyserini.eval.msmarco_passage_eval \
  msmarco-passage-dev-subset \
  runs/run.msmarco-passage.unicoil.tsv
\end{verbatim}
\end{quote}
\noindent The output should be the figure that appears in Table 2 of~\cite{Lin_Ma_arXiv2021}.
Very small differences are sometimes observed due to the inherent non-determinism associated with neural inference (e.g., CPU vs.\ GPU inference, and even across different GPUs).


Let me try to further unpack this ideal of ``two-click reproduction''.
The high-level goal is to reduce friction for users who wish to reproduce a particular result, for example, a figure that is reported in the results table of a paper.
Even if code associated with the paper is available, there's no easy way to separate different phases of the experiment, e.g., training the model from scratch vs.\ inference using a publicly shared model checkpoint.
Even focusing on inference (i.e., ranking), the complexity of modern IR evaluation methodology means that there are a gazillion tiny details to keep track:
Where do I get the index?
Where do I get the topics?
Where do I get the model itself?
What versions of each, exactly?
Is {\it this} the right version that goes with {\it that}?

Sorting through these details is not intellectually challenging, but can be confusing for a novice (e.g., a student just getting into IR) or even a seasoned IR researcher who's never worked with this specific collection before.
As a simple example, there are often different versions of a particular set of queries:\ what everyone calls the 6,980 queries in the MS MARCO passage ranking development set is actually only a subset of the ``real'' full development set.
My other favorite example is TREC-COVID~\citep{RobertsKirk_etal_2020}, which has no less than a dozen different sets of relevance judgments.
All of them are useful, but for answering different research questions.
Which one do you use?

Quite simply, the goal of ``two-click reproduction'' is to relieve the user of all these burdens via a simple, self-contained command that can be copied and pasted into a shell to reproduce an experimental result of interest.
Providing a fully self-contained mechanism with a well-packaged artifact reduces friction; this is about as ``easy to use'' as you can get.

We even provide two-click reproductions for work by others.
For example, the following command reproduces the results of DPR~\citep{karpukhin-etal-2020-dense} on the Natural Questions (NQ) dataset:
\begin{quote}
\begin{verbatim}
python -m pyserini.search.faiss \
  --topics dpr-nq-test \
  --index wikipedia-dpr-multi-bf \
  --encoded-queries dpr_multi-nq-test \
  --output runs/run.dpr.nq-test.multi.bf.trec \
  --batch-size 36 --threads 12
\end{verbatim}
\end{quote}
\noindent See~\citet{Ma_etal_ECIR2022} for additional explorations.
Perhaps the best testament to our efforts is that Pyserini is referenced by~\citet{karpukhin-etal-2020-dense} in their official repo\footnote{\url{https://github.com/facebookresearch/DPR}}
 as the preferred implementation to replicate their work.


The software artifact should have a clearly defined scope, in terms of what it does and, just as importantly, what it doesn't do.
In this case, Pyserini allows a researcher to reproduce a run on the MS MARCO passage corpus using a specific ranking model.
Retrieval is performed using a specific model checkpoint:\ {\it training} a model from scratch is out of scope (although we've shared other code to enable reproducible model training).
Retrieval uses a pre-built index:\ building an index from scratch or searching another corpus is also out of scope in this specific instance (although Pyserini does provide tools for indexing and searching arbitrary corpora).
However, this reproduction command does provide generalizability to different queries, since the encoder model can be applied to arbitrary queries for retrieval.

The scope of a reproducibility effort can often be defined in terms of abstractions encoded in software artifacts.
The application of neural networks can be divided into the training of the retrieval models and inference using those models.
Quite explicitly, Pyserini does not provide any code for training neural models; it is focused on neural inference at search time.

Related to the issue of scope is the explicit acknowledgement in this two-click reproduction ideal that the software artifact and the reproduction commands comprise an abstraction barrier.
The contract is simply that, ``if you run this command, you'll be able to reproduce these results.''
No promises are made about the quality of the code behind the scenes, which may be a pile of spaghetti.
This, I believe, is a feature, not a bug.
Internally, it would be desirable that, once the cover is lifted, the internals of the artifact are beautifully engineered, but this should not be a barrier to reproducibility.
I can't count the number of times I've heard something along the lines of ``the code is really messy, I want to clean it up first before I open source it.''
Mentally, that translates in my mind into ``it'll never happen'', and usually I'm right.
It is difficult to tell if a researcher is using this line as an excuse or if it's uttered in good faith.
In the latter case, other priorities usually intervene, and the net effect is the same.
Code never sees the light of day.

The high-level point is that messy code should not be an impediment to reproducibility, as long as the right abstractions are established---in this case, a PyPI artifact.
Of course, clean internal implementations will make the packaging easier, but janky code that generates the correct behavior is much preferred to elegant code that doesn't work or not having any open-source code at all.
With a messy but functional implementation, there exists a starting point for refactoring down the road if so desired.
There are, literally, entire tomes written about best practices for doing so in a sane manner; for example, I recommend the recent book by \citet{MissingREADME} for practical advice and an entry point into this vast literature.

\section{How?}
\label{section:how}

Having covered ``why'' and ``what'', I move on to cover ``how''.
Repeating from the introduction, my approach can be summarized in two high-level bits of advice:
(1) motivate the importance of reproducibility by appealing to self interest instead of altruism, and (2) engineer social processes to promote virtuous cycles and build standardized tools to reduce technical barriers.
The implementation of these two points should be guided by the dogfood principle, or the directive of ``eat your own dog food'', which refers to the colloquialism of using one's own ``product''.
In the context of a research group, it means that members of the group should be actively using the software artifacts developed by the group.
Quite simply, software artifacts that are used tend to become refined over time, or at the very least, bugs get fixed, because otherwise research would grind to a standstill.

My group uses Pyserini, Anserini, and a few other packages we've developed as the foundation for ongoing work.
Many new research ideas build on, hook into, or otherwise depend on Pyserini.
In turn, improved capabilities in Pyserini spur further advances.

To provide an example, the two-click reproductions described in the previous section solve a number of problems for the community, {\it including ourselves} (invoking self interest again).
Specifically, they provide competitive baselines for comparisons and a solid foundation for first-stage retrieval in a multi-stage ranking architecture.
For example, students focused on building better rerankers need not waste time worrying about the proper setup of the first-stage ranker.
They simply follow the prescriptions in our two-click reproductions as the starting point.
Across the research group, this ensures consistency and reduces the possibilities of bugs:\ We can be confident that every reranker implementation is consuming exactly the same set of candidates and thus the comparisons are fair.

More generally, Pyserini makes it easy to run experiments on standard IR test collections; the toolkit handles much of the boilerplate, such as bindings to query sets, relevance judgments, and evaluation scripts.
Many capabilities come for ``free'', for example, general techniques such as rank fusion and pseudo-relevance feedback can be applied with little effort.
This means that (a) the student writes less code (appealing to self interest again) and (b) the veracity of results increases due to greater consistency in experimental design.

Thus, with the dogfood principle, it's clear to see that reproducibility is driven by self interest.
It allows my students and collaborators to more easily build on each other's results and iterate more rapidly, enhancing their productivity and thus leading to more publications.
{\it We} are the primarily beneficiaries, and the community benefits as a nice side effect.

Here's a sketch of how all these ideas ``tie together'', with repeatability as a starting point.
I'll first describe the social processes that I've engineered to promote virtuous cycles and then move on to discuss infrastructure that reduces the technical barriers to reproducibility.

\subsection{Social Processes: From Repeatability to Reproducibility}

As a starting point, a student is motivated to make experiments repeatable, for all the reasons already discussed in Section~\ref{section:why}.
This involves documenting all the steps necessary to produce experimental results, including configuration settings, command-line invocations, etc.
The documentation that describes how to repeat an experiment (written by the student who initially ran the experiment), when shared, becomes what I call a reproducibility guide (or a ``repro guide'' for short).
In many cases, these guides are just markdown files in the \mytt{docs/} directory of the GitHub repository that contains the code.
They contain, at a minimum, the sequence of command-line invocations that are necessary to reproduce a particular set of experimental results, with accompanying descriptions in prose.
The goal is that copying and pasting commands from the guide into a shell should succeed in reproducing the same experimental results (modulo issues like non-determinism in GPU execution).

The final step is to actually get another person to ``try out'' the guide, i.e., follow exactly the prescribed steps and make sure that they work as expected.
Who does this and why would they?\footnote{Faculty can offer carrots or wave sticks; generally, the former is far more effective.}

I have two tools:\ again, appealing to self interest, but augmented this time with reciprocity, and a new trick---providing an onboarding path to new students.

For students who are already actively contributing to shared code, there are multiple incentives.
Assisting with a reproduction gives the student first access to a new feature, one that could potentially serve as the basis of follow-up work.
Additionally, the social pressures of reciprocity can be an effective motivation:\ students are the beneficiaries of previous group members who ``paved the way'' and thus it behooves them to write good documentation to support future students.
Self interest and reciprocity intersect as well, because students know that at some future point in time, I will demand a check on {\it their} reproduction guide.
It's nice to be able to say, ``Please help me out here, since I did the same for you the last time.''

The other appeal is that reproduction guides provide onboarding paths.
For prospective students who wish to become involved in our group's research, performing reproductions offers exposure to our work and an introduction to our codebase.
These exercises are particularly suitable for undergraduates as their first step in learning about research.
Students are quite motivated (out of self interest) and the group benefits from more people checking the quality of our work.

Sometimes, I try out the reproduction guides myself, and this, too, is motivated by self interest.
The typical scenario is documentation written by a graduating student as a final ``wrap up'' of the work.
From experience, I know that if I ever want another student to build on this work, I had better make sure it works, because once a student graduates, broken code becomes much harder to fix.

I can't emphasize how important it is to actually have someone {\it try out} and follow the reproducibility guide, as opposed to just passively reading the document.
A very common scenario:

\begin{quote}
A: ``This doesn't work, I get the following error\ldots''

B: ``Oh, sorry about that, I forget to check in this file.''
\end{quote}

There is simply no substitute for a hands-on attempt to catch bugs, spot missing details, and uncover hidden assumptions.

Many of these reproduction guides are associated with a ``reproduction log'' at the bottom of the page, which contains a record of individuals who have successfully reproduced the results and the commit id of the code version used.
With these reproduction logs, if some functionality breaks, it becomes much easier to debug, by rewinding the code commits back to the previous point where it last worked.

The social processes that I have described promote and sustain a virtuous cycle, and here we have made the jump from repeatability to reproducibility.
Students begin by recognizing the value of ``packaging'' up their research so that they can repeat their own experiments.
These reproduction guides are made publicly accessible, and are independently confirmed to be functional.
Code that works provides a solid foundation to build on---by both the original authors as well as others.
This in turn accelerates experimental iterations and facilitates rapid explorations of novel ideas built using existing components, ultimately leading to greater productivity.
Students see the payoff of reproducibility efforts and are inclined to sustain their contributions.
And around and around we go.

\subsection{Standardized Tools: From Reproducibility to Two-Click Reproductions}
\label{sec:how:tools}

At a high-level, we can divide reproducibility into the social and technical aspects.
I believe the first is much more important, because any tool to support reproducibility will either be ignored or circumvented unless there are social processes to promote their usage.
The previous section focused on exactly these, and that gets us from repeatability to reproducibility.
Here, I discuss tooling to further lower barriers.
If social processes stimulate ``the will'', standardized tools provide ``the way''.

The core idea is to make investments in tooling to automate the process of ``going through'' the reproduction guides.
In Anserini and Pyserini, we have built extensive regression tests with elaborate test harnesses (sometimes called jigs).
These regressions are tightly integrated with the two-click reproductions discussed in the previous section:\ in fact, in many cases, the regression tests simply wrap the execution of the two-click reproduction commands and verify that the outputs match stored specifications.
The periodic execution of these regressions ensures that the two-click reproductions continue to ``work as advertised''.

In Anserini, we have implemented a test harness called 
\mytt{run\_regression.py} that takes as input a YAML configuration file for a set of experimental conditions on a standard IR test collection, for example, the MS MARCO V2 passage test collection~\citep{Craswell_etal_TREC2021,Ma_etal_SIGIR2022}.\footnote{\url{https://github.com/castorini/anserini/blob/master/src/main/resources/regression/msmarco-v2-passage.yaml}}
Here's a sample invocation:
\begin{quote}
\begin{verbatim}python src/main/python/run_regression.py \
  --index --verify --search --regression msmarco-v2-passage    
\end{verbatim}
\end{quote}
\noindent The script runs through the following steps:\ It builds the index from scratch (i.e., the raw corpus), verifies index statistics (e.g., number of documents and terms processed), performs retrieval runs using different retrieval models (e.g., BM25 with Rocchio feedback), evaluates the outputs (e.g., with \mytt{trec\_eval}), and checks effectiveness figures against expected results (stored in the configuration).

That is, the execution of this script verifies the reproducibility of a set of experimental conditions in a fully automatic manner.
Upon each successful execution, the regression script generates a documentation page from an associated template, populating the results (e.g., average precision) from the trial.\footnote{\url{https://github.com/castorini/anserini/blob/master/docs/regressions-msmarco-v2-passage.md}}
All of this happens without any human intervention.
The script depends on having access to the raw corpus, which is stored on our group's servers at known file system locations.
However, the corpus path is a configurable parameter, so anyone can run the same regression test if they have a copy of the corpus.

There are currently around three hundred such tests, which take several days to run end to end (orchestrated by yet another script).
The largest of these builds a 10 TB index on all 733 million pages of the ClueWeb12 collection.
Although it is not practical to run these regression tests for each code change, we try to run them as often as possible, resources permitting, to catch new commits that break existing functionalities early so they are easier to debug.
These regression tests are always run before a formal release of the toolkit, prior to publishing an artifact on Maven central, to ensure that released jars produce the expected experimental results.

On top of the regression framework in Anserini, further tests in Pyserini compare its output against Anserini's output to verify that the Python interface does not introduce any bugs.
These are written as Python unit tests and, for example, check different parameter settings from the command line, ensure that single-threaded and multi-threaded execution yield identical results, that pre-built indexes can be successfully downloaded.

\begin{figure*}[t]
\centering
\includegraphics[width=0.85\textwidth]{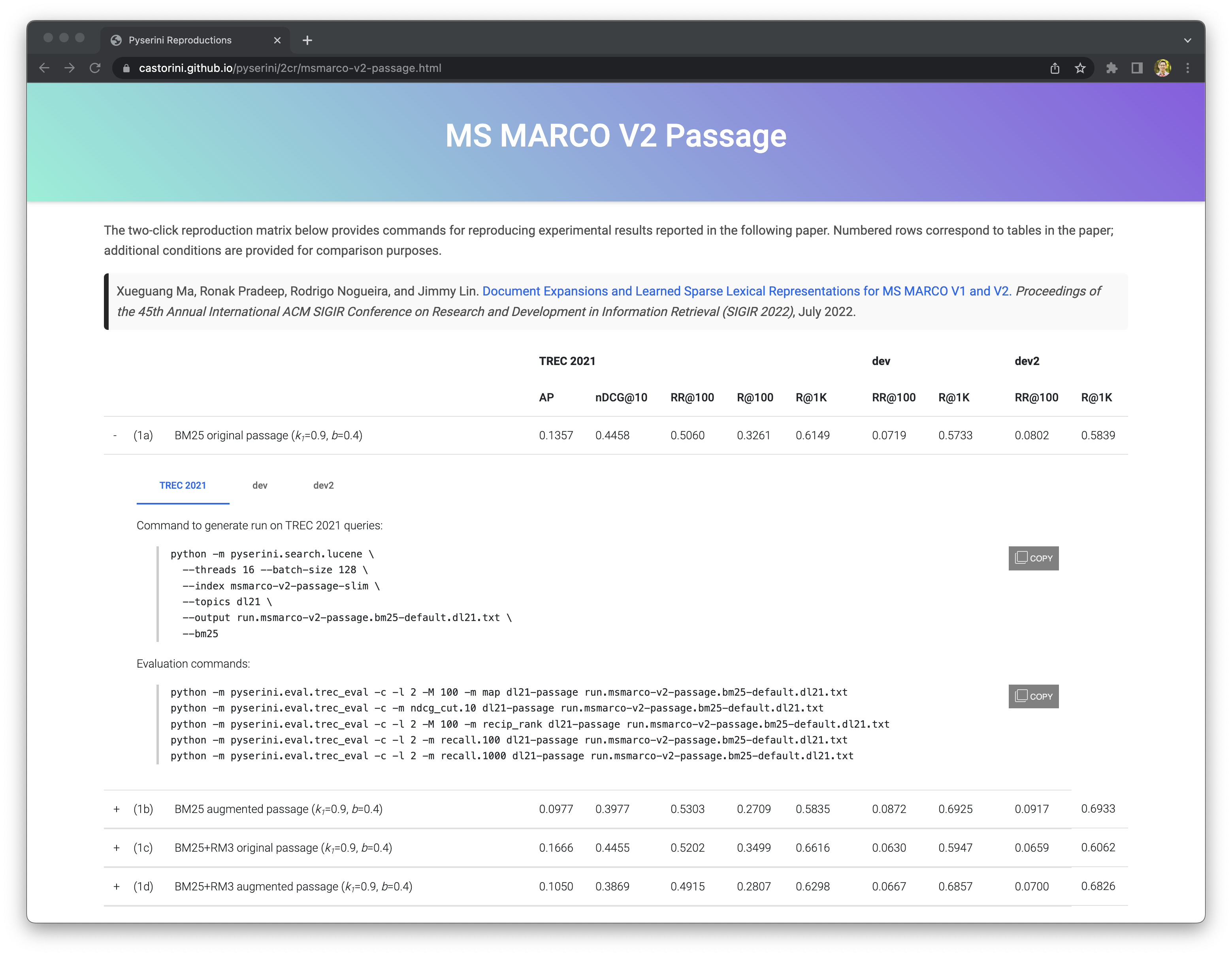}
\caption{The reproduction matrix for the MS MARCO V2 passage corpus, which is available at \url{https://castorini.github.io/pyserini/2cr/msmarco-v2-passage.html}. Each row represents an experimental condition and is associated with two-click reproduction commands.}
\label{fig:2cr}
\end{figure*}

In Pyserini, experimental conditions are gathered together and organized into what we call a reproduction matrix, an example of which is shown in Figure~\ref{fig:2cr} for the MS MARCO V2 passage corpus.
Each row illustrates a particular experimental condition, while the columns show evaluation metrics with respect to different sets of queries.
A row can be expanded to reveal the commands necessary to generate those evaluation figures.
These are exactly the two-click reproductions described in Section~\ref{section:what}, organized in an easy-to-consume format.

Finally, the reproduction matrix is backed by another script that programmatically iterates through all rows (experimental conditions), performs retrieval with the specified invocations, and verifies that the evaluation scores are as expected (i.e., checks each cell in the table).
In this case, the command is:
\begin{quote}
\begin{verbatim}
python scripts/repro_matrix/run_all_msmarco.py --collection v2-passage
\end{verbatim}
\end{quote}
\noindent In fact, the reproduction matrix webpage is automatically generated by the above script upon successful completion.
Once again, these regression experiments are quite computationally expensive, and collectively they take several days to run.
Nevertheless, these checks are performed before every artifact release on the Python Package Index.
Thus, we are confident of the reproducibility of various retrieval models implemented in Pyserini, to the extent that they are covered by these tests.

As I've previously argued, reproducibility is a continual process, not a ``one and done'' deal~\citep{Lin_Zhang_ECIR2020}.
The testing infrastructure described here ensures that our two-click reproductions continue to work even as the entire codebase evolves and gains new features.
What I've described here can be characterized as a custom continuous integration/continuous delivery (CI/CD) framework, adapted to the unique characteristics of research.

This might all sound like a lot of work to set up initially, and indeed it was.
However, all the upfront engineering costs have already been ``paid for''.
For a student building a test case for a new experimental condition, the effort is relatively modest, and the process consists mainly of writing configuration files and hooking into the test infrastructure.
Once connected, the student can be confident that the code will continue to generate the expected retrieval results.
Down the road, when it is time to write up the thesis, there's no need to ``dust off'' the code to make sure it ``still works''.
The regressions tests ensure that it never stopped working.

In summary, reproducibility has become ingrained as a shared norm in our group, operationalized in social processes and facilitated by technical infrastructure.
I think that this has allowed us to nicely balance the demands of reproducibility with the ability to iterate rapidly.

\section{Other Considerations}
\label{section:other}

As promised in the introduction, this section discusses a smörgåsbord of issues that don't fit neatly into the ``what'', ``why'', and ``how'' narrative.

\subsection{Scoping and Timing of Reproducibility Efforts}

In truth, written reproduction guides and automated regression testing lie along a spectrum of ``reproduction rigor'', with different cost/benefit tradeoffs.
Although we aim to have reproduction guides for every paper, only a relatively small fraction of our group's research becomes ``enshrined'' in automated regressions that are maintained over time.

We currently do not have clear-cut criteria as to which retrieval models or experimental results receive the regression treatment, but as a rough heuristic, we use the following question as a guide:\ Is this work we'd like to extend further?
If so, then we would go about building an appropriate regression.
Similarly, if we find a paper by others that we'd like to build on (as in the case of DPR), we would make the investment to replicate the work and to build regressions into our codebase.

As already mentioned in Section~\ref{section:what}, scoping the effort is an important part of the reproducibility discussion.
Consider the common case of a modeling advance that is described in a paper, i.e., we proposed a novel retrieval model that appears to be better than previous work, and the contribution represents a fruitful line of inquiry that the group hopes to push further.
In this case, building an appropriate regression makes sense.
However, to balance cost and reward, we do {\it not} construct regression tests for every experimental result reported in the paper.
Instead, we are guided by the question:\ In a follow-up paper, which of the existing experimental conditions from this paper would serve as the baseline for comparison?
{\it That} becomes the target for integration into our regression framework.
We find that building tests for ineffective contrastive settings or ablation conditions provides little value relative to the amount of effort required.

Part of the scoping exercise is to determine what aspects of the proposed model should be included in which codebase.
If the original experiments were performed with Pyserini to begin with, then the answer is straightforward:\ Model checkpoints are made public (e.g., on the Huggingface model hub) and two-click reproductions are directly integrated into Pyserini.
However, since our toolkit (by design) does not include code for model training, reproduction guides for that aspect of the work must go elsewhere, typically in another code repository.

In some cases, a novel retrieval model does not neatly fit into the design of Pyserini.
These model implementations (both training and inference) usually begin their lives in a separate repository, but as part of the reproducibility planning exercise, we debate whether it is worthwhile to import the model inference code into Pyserini so that end-to-end retrieval experiments can be conducted alongside all the other available models in a seamless manner (as part of a reproduction matrix, see Section~\ref{sec:how:tools}).
These decisions are made on a case-by-case basis.

It is worth explicitly noting that any inclusion to the constantly growing test suites in Pyserini and Anserini represents an open-ended maintenance commitment for the life of the project.
Any addition, in essence, incurs a permanent liability on the group (and as I'll discuss in Section~\ref{section:misc:leader}, this burden usually falls on me).
There's no point in adding a model to the regression framework unless there's the intention of keeping the code functional in the long term.
Once added, we almost never abandon a regression test, except in very rare circumstances, for example, a failure due to changes in underlying code that we depend on but have no control over.

Operationally, continual expansion of test suites means that the complete set of regressions takes longer and longer to run, which has the practical effect of slowing down release iterations (e.g., on PyPI).
However, I don't think this has impacted the iteration speed of individual students since components in the codebase are largely decoupled.
Nevertheless, servers continue to get faster and more powerful, so I think our current operations are sustainable.

I've found that the best time to make investments in long-term reproducibility is the window between the acceptance notification of a paper and the final camera-ready deadline.
This provides an opportunity to perform a ``final check'' on the results and to plan for the long-term maintenance of the model.
Work during this time window also ensures that the evaluation results reported in the final paper version match the figures that can be recreated with our two-click reproductions.

In some cases, the journey from reproduction guides to automated tests is circuitous.
For example, we might not have found a particular thread of work sufficiently promising to have integrated it into our regression framework, but subsequent developments changed our minds.
It is never too late, but I have encountered cases of ``reproducibility debt'', much like the notion of ``technical debt'' in software engineering.
The complexities of modern software stacks create hidden dependencies that often break retrieval models in subtle ways as code evolves.
Especially if someone has not tried out a reproduction guide in a while, it might be discovered later that the results have changed.
Repeatability is a fickle beast.

\subsection{Bootstrapping Reproducibility}

What about the cold start process?
The foregoing discussions describe the operational aspects of reproducibility in my group in ``steady state'', where virtuous cycles have already been established.
Existing software artifacts are already functional and the benefits of using them are evident.
Processes and shared norms are in place, and tools to simplify the routine have been built.
Once again, motivated by self interest, the value that can be extracted by participating in, for example, our Pyserini reproducibility ecosystem is greater than the costs.

What if this isn't the case?
How can a research group start the flywheel spinning from a standstill?\footnote{To use an analogy attributed to Jeff Bezos: Virtuous cycles are like flywheels; they hold a lot of energy and are difficult to slow down, but they're even harder to spin up initially.}
Before addressing this point, it is important to recognize that the reproducibility narrative I've articulated here does not work for everyone.
Only certain ``styles'' of systems-oriented research organized around software artifacts are conducive to the treatment described in this essay.
However, to anyone who wishes to replicate a similar culture of reproducibility:

I admit that getting the flywheel spinning is hard, and the truthful answer is:\ I don't really know how, at least in a replicable manner.
I began my academic career as an assistant professor in 2004 and have started countless research projects that involve building and sharing software artifacts.
Only recently have I successfully pulled together the elements that sustain a culture of reproducibility.
Of course, I could construct a story of our success, but it would merely be a post-hoc narrative.
The casual factors are too complex and the training examples are too few to build an explanatory model.

Nevertheless, I will share some ideas that are independently worthwhile, regardless of their actual contributions to reproducibility.
First, adopt software engineering best practices.
A research environment is of course different from a production shop geared towards delivering products and services to customers, but this doesn't mean that research groups should ignore mature and well-established processes.
Pyserini and Anserini both adopt standard best practices in open-source software development.
The code is available on GitHub, issues are used to describe proposed feature enhancements and bugs, and code changes are mediated via pull requests that are code reviewed.

Second, look for opportunities where a long-term research agenda aligns with the construction of software artifacts that can be generalized into a toolkit, library, or platform.
Reproducibility efforts have substantial upfront costs that need to be amortized over several years to achieve a net gain.
The (planned) software toolkit should ideally provide the basis for several related research projects that yield multiple publications.
Without this long-term vision and commitment to a shared codebase, the group might never reap the rewards of the initial investment.
With a plan in place, it is possible to make progress incrementally.
For example, the multi-layered regression frameworks in Anserini and Pyserini evolved over many years.
However, the commitment to build a toolkit for tackling the core ranking problem in information retrieval was made on day one.

How does one identify these opportunities?
For junior faculty, their own research statements provide the inspiration!
As an integral part of the application process for academic positions, the research statement should contain a coherent, multi-year research agenda.
Look there for possible alignment, as the vision should have already been articulated clearly.

Third, the richness of the modern software ecosystem means that opportunities for contributing software artifacts can happen at many different layers in the stack, and specialized niches abound.
For example, Pyserini relies on PyTorch and Anserini relies on Lucene.
It'd make little sense for our group to try and build the equivalent of either PyTorch or Lucene from scratch.
Similarly, it might not make sense for another research group to build an independent IR toolkit from scratch.
Instead, join us and build on top of our toolkits to handle a niche that is not currently well served.
We'd welcome your contributions!

Finally, leadership is critical and deserves a dedicated section.
I turn to this next.

\subsection{The Critical Role of Leadership}
\label{section:misc:leader}

I am the overall architect of Pyserini and Anserini.
I am the person who usually runs the regression tests, shepherds the release of software artifacts, and generally keeps tabs on everything that is going on.
In software development terms, I am not only the engineering manager but also the tech lead.
This makes sense from multiple perspectives, as I am in the best position to serve as the long-term institutional memory of the research group.
Students come and go, but my presence (hopefully) remains constant.
Many of the retrieval models in Pyserini were built before the arrival of my most recent cohort of students, and some models will continue to be refined even after they graduate.
I have the most complete view of what everyone is working on, and this allows me to coordinate multiple overlapping research projects.

For example, I regularly introduce students to existing features in Anserini and Pyserini that can expedite their research.
In many cases, showing students how to use a model is as simple as pointing them to the corresponding reproduction guide and asking them to go through it, or even better, directing them to the documentation that provides the two-click reproduction commands.
Another common pattern is that I arrange ``handoffs'' from a graduating student to a new student who wishes to continue pursing a related line of work.
If the practices described here are faithfully executed, this is a relatively seamless process.

It is important for the group leader to assume the roles described above---in simple terms, serving as {\it both} the engineering manager and the tech lead.
Having this mindset in my opinion is one key to sustaining a culture of reproducibility.
For example, I have written most of the test harness code in Pyserini and Anserini, and in many cases, I end up writing unit tests for students.
This can be characterized as a ``servant leadership'' style:\ writing testing frameworks certainly isn't glamorous, but it's critically important.
Working on these bits of code is the best use of my time from the perspective of benefiting the entire group---as investments in reproducibility pay dividends for everyone using the codebase---and fulfills my personal desire to stay technically engaged with students.

Starting an academic research group has been analogized to running a startup, with the faculty member as the CEO.
In the context of the North American academic system, this analogy is apt, as faculty members generally lead their own groups.
In the beginning, they must do everything, including assuming the roles of an engineering manager (e.g., hiring and mentoring students) as well as the tech lead (e.g., guiding students' technical progress and examining their implementations).

However, as a faculty rises through the academic ranks and grows a research group, a common organizational pattern is to cede the role of the tech lead to a senior graduate student.
While certainly workable, this comes with associated risks.
Students (eventually) graduate (hopefully!)\ and the next stage of their careers may no longer have room for the role.
Arranging for succession ``handoffs'' become more difficult if the group leader is not intimately involved.
This is not unlike what happens in a corporate environment:\ When a tech lead departs, management needs to find a replacement, typically elevating another member from the same project.
In the academic context, this is likely another senior graduate student working on similar research problems, provided one exists.
Needless to say, sustaining research momentum is much easier if there is continuity, as in the case where the group leader also serves as the tech lead.

\section{Conclusions}

In the end, building a culture of reproducibility is hard work, but the rewards are worthwhile.
Coming back to where I started:
I am passionate about making research reproducible by building and sharing software artifacts that the community can use to recreate work by us and other researchers.

One of the most flattering comments I've ever received from a colleague reflects exactly this passion:
\begin{quote}
One of the most valuable contributions of Jimmy's work is the way he shares everything you need to reproduce it. He figures something out and he *shows* everyone how to do it. I cannot applaud his efforts on this front enough. It is enormously valuable to our community. I constantly learn things by looking at what Jimmy does -- not just descriptions in writeups. I am a researcher because I love learning how things work. There are few people in our community doing a better job at teaching people how to learn what he has learned than Jimmy.
\end{quote}
Some of the best comments I've recently received from anonymous reviewers are in the same vein. (To those reviewers:\ Thank you!)
The academic community has begun to recognize the importance of reproducibility and to reward such efforts with dedicated tracks and associated publications in top-tier venues.
I think we're heading in the right direction, and I hope that this essay offers some insights on how we can continue building a culture of reproducibility everywhere.

\section*{Acknowledgements}

None of the work described here would have been possible without my students and other members of my research group, who have not only contributed to our group's many open-source software artifacts, but have also served as guinea pigs to the social processes that I have tried to engineer.
The other crucial ingredient, of course, is funding, and I gratefully acknowledge support primarily from the Natural Sciences and Engineering Research Council (NSERC) of Canada and the Canada First Research Excellence Fund, with computational resources provided in part by Compute Ontario and Compute Canada.

\bibliographystyle{ACM-Reference-Format}
\bibliography{repro}

\end{document}